\documentclass{solarphysics}
\usepackage[hyperref,optionalrh,solaromanenum]{spr-sola-addons} 
\usepackage{epsfig}                     
\usepackage{graphicx}                    
\usepackage{amssymb}
\usepackage{amssymb}                    
\usepackage{color}                       
\usepackage{url}
\usepackage{breakurl}                         
\usepackage{natbib}

\newcommand{\etal}{et~al. }

\newcommand{\caiih}{Ca\,{\sc ii}\,{\sc H}}

\newcommand{\heii}{{He\,{\sc ii}\,}}

\newcommand{\mgvi}{{Mg\,{\sc vi}\,}}
\newcommand{\mgvii}{{Mg\,{\sc vii}\,}}
\newcommand{\feviii}{{Fe\,{\sc viii}\,}}
\newcommand{\fexii}{{Fe\,{\sc xii}\,}}

\newcommand{\caxvii}{{Ca\,{\sc xvii}\,}}

\newcommand{\ov}{{O\,{\sc v}\,}}
\newcommand{\sivii}{{Si\,{\sc vii}\,}}

\newcommand{\Ha}{${\rm H\alpha}$ }
\newcommand{\ha}{${\rm H\alpha}$ }

\def\arcsec{\hbox{$^{\prime\prime}$}}

\begin{document}

\begin{article}

\begin{opening}

\title{Probing the quiet solar atmosphere from the photosphere to the corona}

%
\author{Ioannis~\surname{Kontogiannis}$^{1}$\sep
        Costis~\surname{Gontikakis}$^{1}$\sep
        Georgia~\surname{Tsiropoula}$^{2}$\sep
        Kostas~\surname{Tziotziou}$^{2}$\sep
          }
%
\runningauthor{Kontogiannis \etal}
\runningtitle{Quiet solar atmosphere}

 \institute{$^{1}$ Research Center for Astronomy and Applied Mathematics (RCAAM) Academy of Athens, 4 Soranou Efesiou Street, Athens, GR-11527, Greece
                     email: \url{jkonto@noa.gr} \\
             $^{2}$  Institute for Astronomy, Astrophysics, Space Applications and Remote Sensing, National Observatory of Athens, GR-15236, Penteli, Greece
                     email: \url{georgia@noa.gr} email: \\}
\begin{abstract}
We investigate the morphology and temporal variability of a quiet Sun network region in different solar layers. The emission in several EUV spectral lines through both raster and slot time series, recorded by EIS/Hinode is studied along with H$\alpha$ observations and high resolution spectropolarimetric observations of the photospheric magnetic field. The photospheric magnetic field is extrapolated up to the corona showing a multitude of large and small scale structures. We show for the first time that the smallest magnetic structures both at the network and the internetwork contribute significantly to the emission in EUV lines, with temperatures ranging from 8$\cdot$10$^{4}$\,K to 6$\cdot$10$^{5}$\,K. Two components of transition region emission are present, one associated with small-scale loops that do not reach coronal temperatures and another one acting as an interface between coronal and chromospheric plasma. Both components are associated with persistent chromospheric structures. The temporal variability of the EUV intensity at the network region is also associated with chromospheric motions, pointing to a connection between transition region and chromospheric features. Intensity enhancements in the EUV transition region lines are preferentially produced by H$\alpha$ upflows. Examination of two individual chromospheric jets shows that their evolution is associated with intensity variations in transition region and coronal temperatures. 
\end{abstract}
%

\end{opening}

%
\section{Introduction}
\label{s:intro}

The quiet solar atmosphere is highly variable in both space and time, due to the interaction between magnetic fields, solar differential rotation and convective flows. At the solar surface, the large scale convective flows form a pattern of cells known as supergranular cells. The magnetic field lines are transported horizontally at the edges of the supergranular cells, resulting to the formation of the magnetic network, a web-like pattern, bright in strong resonance lines, such as Ca\,{\sc ii}\,{\sc H}, that encompasses the slightly darker internetwork (IN). At the photosphere, the network consists of bright points that represent the cross-sections of magnetic flux tubes \citep{schrijver97}. At the chromosphere, where the magnetic forces gain full control of the plasma dynamics, numerous elongated structures, seen in absorption in H$\alpha$, stem from the network and cover part of the IN. These are traditionally called mottles \citep[see][for a recent review]{tsirop12} and are believed to outline the magnetic field of the chromosphere. The network pattern ``survives'' up to coronal temperatures \citep{gont2003,tian08c,tian08b}, while at the corona, large scale closed magnetic loops and/or magnetic funnels dominate \citep{peter01}.
\par
The transition region (TR), that lies between the hot and tenuous solar corona and the chromosphere, was modelled by \citet{gab76}, who took into account the magnetic field geometry resulting from the supergranular flows and the energy balance. In this model, the magnetic field flux tubes expand rapidly with height and form funnels that fill the TR and corona, while all energy necessary to maintain the TR was considered to be provided by heat conduction flowing downward from the corona. However, although this model gives consistent results for the differential emission measure (DEM) for temperatures higher than 10$^{5}$\,K, it fails to explain the observed DEM at lower temperatures. 
\par
\citet{dowdy86} suggested that small loops, confined within the network boundaries, with relatively low temperatures might contribute a significant amount to the lower TR emission. Since these loops are disconnected from the corona, they must be heated by an internal process. \citet{feldman83} presented observational evidence that the TR plasma in the temperature range 4$\cdot$10$^{4}$--2.2$\cdot$10$^{5}$\,K originates from the sum of unresolved, magnetically and thermally isolated structures, which he called ``unresolved fine structures'' (UFS). Spectroscopic observations bear signatures of the coexistence of loops with various sizes at the TR \citep{peter01}. In fact, a significant part of the magnetic flux of the network may be connected to nearby internetwork magnetic elements, forming low lying closed loops that surround the network \citep{schrijver03}. Recent high resolution observations show that cool loop-like structures, thermally isolated from the corona, may exist in the TR quiet Sun network \citep{patsourakos07,vourlidas10}. \citet{hansteen14} showed that small-scale structures (or the so-called UFS) are abundant in the TR, at least in lines formed at 10$^{5}$\,K plasma, imaged by the Interface Region Imaging Spectrometer (IRIS) mission \citep{iris}.
\par
From a theoretical point of view, a mixture of small cool and intermediate-temperature loops with temperatures between 10$^{5}$ and 10$^{6}$\,K could reproduce the observed emission in the lower TR \citep{sasso12,sasso15}. State-of-the-art simulations \citep{guerreiro13} show that the Si\,{\sc iv}\,1393\,{\AA} TR emission comes in fact from cool, low-lying loops. \citet{schmit16} came to the same conclusion by combining IRIS observations and simulations. Their conclusions were partly based on detailed statistics between relatively low-resolution observations of the magnetic field (for QS standards) and radiance in the relatively cool (for TR standards) Si\,{\sc iv}\,1393\,{\AA} ($logT=4.9$) emission line. Therefore, studies that combine emission in both cool and hot emission lines can shed light to the existence of the different types of loops (in height and temperature). Even more, such studies, when combined with high resolution magnetograms can help us understand the magnetic connectivity between the different solar layers.
\par
Regarding the temporal variability of the TR emission, until recently, localized broadenings of the TR spectral lines during short periods of time \citep[explosive events;][]{Brueckner_bartoe83} and intensity enhancements with lifetimes of the order of a few minutes \citep[blinkers;][]{harrison97,bewsher03} were intensely studied. In retrospect, it is reasonable to assume that past resolution limitations permitted the detection only of the largest/brightest (in terms of spatial extent/intensity) of the TR variations. Indeed, higher spatial resolution observations of the TR revealed that small-scale, jet-like features are highly abundant \citep{tian14}. It has also been found that small-scale chromospheric jets that exhibit rapid Doppler shift excursions \citep{rouppe09} are associated with plasma up to 8$\cdot$10$^{4}$\,K \citep{rouppe15}, which is interpreted as a sign of their heating to TR temperatures. A statistical study by \citet{henriques16}, showed that rapid chromospheric events are partly associated with low-corona (10$^{6}$\,K) brightenings.
\par
The possible association of chromospheric features with motions in the upper atmosphere is not new. It has been established that mottles and spicules carry more than enough mass to load the corona and the solar wind and that the major fraction of this flux returns back to the chromosphere \citep{tsirop_tzio04}. In addition, these authors suggested that this returning flux could explain the persistent redshifts observed in the TR \citep{peter_judge99,gont_dara2000,tian10}. The possible recycling of the solar plasma from the chromosphere to the corona and back has also been demonstrated with EUV spectroscopy \citep{mcintosh_depontieu09b,mcintosh_depontieu09a}.
\par
In this study we address these issues using a unique dataset, which covers a remarkable range of atmospheric layers of a quiet solar region. We investigate the morphology and dynamics of different atmospheric layers/temperature regimes in relation to the magnetic field and explore the imprint of fine-scale chromospheric phenomena in the low-signal EUV measurements.

\begin{table}[!ht]
\caption{EIS Transition region and coronal spectral lines and XRT filter formation temperatures.}
\label{filters}
\begin{tabular}{c c}
\hline
\hline
Spectral line/filter & \textit{T} (K) \\
\hline
\heii 256.32\,{\AA} & $10^{4.9}$ \\
\ov 192.80\,{\AA} & $10^{5.4}$ \\
\feviii 185.21\,{\AA} & $10^{5.7}$ \\
\mgvi\ 269.00\,{\AA} & $10^{5.7}$ \\
\mgvi\ 270.40\,{\AA} & $10^{5.7}$ \\
\mgvii\ 278.39\,{\AA} & $10^{5.8}$ \\
\mgvii\ 280.75\,{\AA} & $10^{5.8}$ \\
\sivii\ 275.35\,{\AA} & $10^{5.8}$ \\
\fexii\ 195.12\,{\AA} & $10^{6.2}$ \\
``C\_poly''            &  $10^{6.2}$ \\
\hline
\end{tabular}
\label{Table:eis_lines}
\end{table}

\section{Data and Analysis}
\label{s:data_analysis}

\begin{figure}[ht]
\centerline{\includegraphics[width=1.05\textwidth]{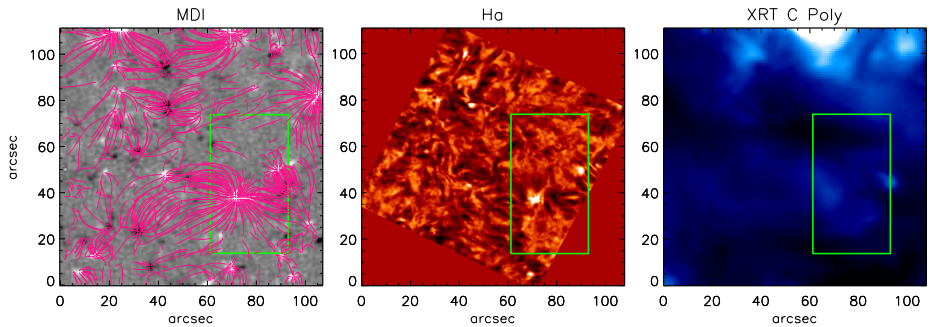}}
    \caption{Temporally averaged images of a quiet Sun region at the disk center, obtained in October 15, 2007. From \textit{left} to \textit{right}: MDI magnetogram overlaid with the magnetic field lines, \ha line center filtergram and XRT-``C\_poly''. The \textit{green} boxes indicate the common EIS and SOT/SP FOV.}
    {\label{fig:fig1}}
\end{figure}

On October 15, 2007, several ground--based and space borne instruments were coordinated to observe the same quiet region located at the solar disk center (Figure~\ref{fig:fig1}). In this work we utilize data from that campaign, and more specifically those obtained with the \textit{Solar Optical Telescope} and its Spectropolarimeter \citep[SOT/SP][]{sot}, the \textit{EUV Imaging Spectrometer} \citep[EIS;][]{eis} and the \textit{X-Ray Telescope} \citep[XRT;][]{xrt}, onboard \textit{Hinode}, the \textit{Michelson Doppler Imager} \citep[MDI;][]{mdi} onboard the \textit{Solar and Heliospheric Observatory} (SoHO), as well as \ha observations obtained with the ground-based \textit{Dutch Open Telescope} \citep[DOT;][]{rutten2004}.
\par

MDI provided high-resolution magnetograms of the region between 08:00--11:00\,UT (Figure~\ref{fig:fig1}, left panel). The cadence of the time series is 60\,s and the spatial scale is 0.6$\arcsec$. The SOT/SP performed two raster scans of the same area in the 6301.5 and 6302.5\,{\AA} Fe\,{\sc i} lines at 09:05 and 09:15 UT. The field-of-view (FOV) of the raster scan (centered on the solar disk centre) was 50$\arcsec\,\times\,164\arcsec$ and the spatial scale of the scans 0.32$\arcsec$. The HAO/CSAC\footnote{\url{http://www.csac.hao.ucar.edu/csac/dataHostSearch.jsp}} team produced the corresponding photospheric vector magnetogram by performing the inversion of Stokes spectra via the MERLIN code. The line-of-sight (LOS) component of the magnetic field from MDI and SOT/SP was then used as an input to a current-free (potential) field extrapolation \citep{schmidt64} to calculate the vector of the magnetic field up to 10\,Mm. Further details on the inversion of the Stokes spectra and the comparison between MDI and SOT/SP data and extrapolated magnetic fields can be found in \citet{kontogiannis11}.

The EIS images plasma with upper chromosphere up to coronal temperatures. The EIS data include a series of 40$\arcsec$ slot images, recorded between 08:14:40 and 09:40:15~UT with a 64\,s cadence and a raster scan taken between 08:51-09:32~UT, with the 2$\arcsec$ slit and 1\,min exposure time. Out of the ten recorded spectral windows nine spectral lines were utilized, which are presented in Table~\ref{Table:eis_lines}, along with the corresponding ion formation temperature, taken from the CHIANTI database (v7.0). 
\par
Most of the available spectral lines are blended. These are either self-blends and/or lines with very high temperatures and very low intensities. For most of the lines, a single Gaussian was used to perform the fits. The \mgvii 280.75/278.39\,{\AA} and \mgvii 278.39\,{\AA}/\mgvi\,269.00\,{\AA} line ratios are used to measure the electron density and temperature at the TR \citep{young07}. The \caxvii spectral line complex was treated following the method described by \citet{levens15}, in order to calculate the intensity and velocity of the \ov line. Given the unknown influence of blends in the width measurements, we only focus on the extracted peak intensity and velocity values.  
\par
The XRT provided time series of ``C-poly'' filtergrams of the solar disk center, with 25\,s cadence and 1$\arcsec$ spatial scale (Figure~\ref{fig:fig1}, right panel). The temperature response of ``C-Poly'' peaks at logT = 6.2. The acquisition time is not regular because the time series is often interrupted to obtain full-disk filtergrams. These filtergrams were removed and were not considered in the present analysis.
\par
DOT observed the same region between 08:32--09:53\,UT, providing speckle-reconstructed images in five positions along the \Ha profile (line centre, $\pm$0.35\,{\AA} and $\pm$0.70\,{\AA}). The spatial scale of each image is 0.109\,arcsec/pixel and the cadence is 30\,s. Further details on preliminary reduction of the DOT observations can be found in \citet{rutten2004}. DOT data were rotated to match the FOV of the space-borne instruments (Figure~\ref{fig:fig1}, middle panel). Using the intensities at opposite positions $\Delta\lambda$ of the line profile at $\pm$0.70\,{\AA}, the Doppler Signal (DS) was calculated at every pixel of the FOV through the formula:

\begin{equation}
{ DS=\frac{I(+\Delta\lambda)-I(-\Delta\lambda)}{I(+\Delta\lambda)+I(-\Delta\lambda)}\;\;\;.    }
 \label{eq:1}
\end{equation}

\noindent and maps of DS were constructed. Positive (negative) DS denotes upward (downward) motion and is a qualitative representation of velocity \citep{tsirop00}.
\par

The co-alignement of observations from different instruments, representing very different atmospheric regimes is usually a challenging procedure and requires a substantial amount of time and effort. First, we co-aligned the blue wing of \ha with the co-temporal \caiih\ images taken by the SOT (not used in this study) by matching the network bright points, visible in the blue wing of H$\alpha$ \citep{leenaarts06}, with the ones in Ca\,{\sc ii}\,{\sc H} and by using cross-correlation for fine-tuning the results. In a similar way, the MDI and SOT/SP magnetograms were co-aligned with H$\alpha$, while the SOT \caiih\ image was co-aligned with the co-temporal EIS \heii slot images. The latter show the same network structure of the chromosphere but the boundaries are more extended. For this reason, the higher-resolution \caiih\ images were first degraded to resemble the \heii images. The XRT images were co-aligned with the \fexii\ EIS slot images. Between the two wavelength ranges of EIS, there is a systematic offset (18\arcsec\ and 1\arcsec\ in the y- and x-direction, respectively) which was corrected by shifting the EIS short-wavelength windows. The EIS slot images were co-aligned with the raster intensity maps taken at the same wavelengths. We estimate that the common FOV is co-aligned with sub-arcsec accuracy.  

\section{Results}
\label{s:results}
\subsection{Quiet Sun structure from the photosphere to the corona}
\label{s:qs_struct}
\subsubsection{Morphology}
\label{s:morph}

In this section we infer the association between the emission in EUV channels, the underlying chromosphere and the overlying corona. We examine the spatial distribution of the emission/absorption in different wavelengths and compare it with the extrapolated chromospheric and coronal magnetic field.

Figure~\ref{fig:fig1} provides the context of the observations by showing a region wider than the region-of-interest (ROI, marked by the green boxes, which outline the common FOV of all instruments). The MDI magnetogram at the left panel reveals the positive and negative magnetic patches that form the network. The field lines of the potential magnetic field are overplotted to show the connections between magnetic concentrations. These connections are also partly outlined by the dark elongated mottles \citep{tsirop12}, seen in the \ha line centre filtergram, in the middle panel of Figure~\ref{fig:fig1}. These low-$\beta$ structures form the magnetic canopy \citep{kont10b} and obscure most of the network boundaries (appearing bright in H$\alpha$ line wings) and IN. The XRT image (Figure~\ref{fig:fig1}, right panel) shows the overlying corona.

\begin{figure}[htp]
\centerline{ \includegraphics[width=1.05\textwidth]{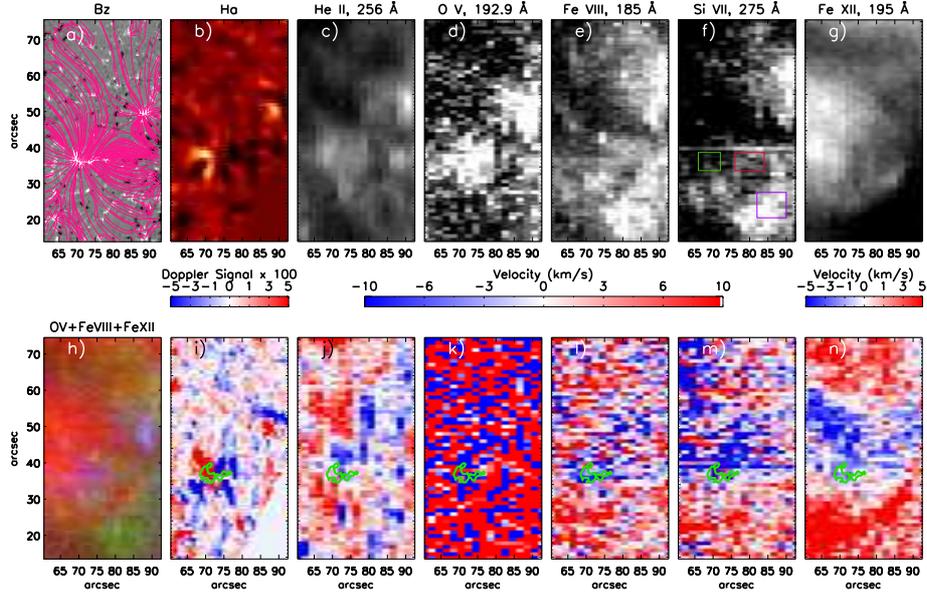}}
  \caption{a) SOT/SP magnetogram with overlaid magnetic field lines, b) \ha line centre intensity synthetic raster and c -- g) peak intensities in \heii 256.32\,{\AA}, \ov 192.80\,{\AA}, \feviii 185.21\,{\AA}, \sivii 275.35\,{\AA} and \fexii 195.12\,{\AA}, obtained from the EIS raster scans. The \textit{rectangles} in the \sivii intensity map denote the areas where TR temperature and density were measured (see text). h) RGB image showing emission in \ov\ (\textit{blue}), \feviii\ (\textit{green}) and \fexii\ (\textit{red}).  i) \ha DS$\pm$0.70\,{\AA} synthetic raster, and j -- n) Doppler velocities calculated from the EIS raster images of the \textit{top} row. Here, \textit{blue} (negative) corresponds to upward motion (towards the observer). The green contours mark the location of the network boundary. All images are cut-outs that correspond to the common FOV marked in Figure~\ref{fig:fig1}. The solar disk centre is located at $\sim$(65,30). }
    {\label{fig:rasters}}
\end{figure}

The positive polarity patch at (x,y)$\sim$(70,35) is part of this network. It is simultaneously observed by all instruments and it is the focus of this study, along with its nearby IN. It connects to opposite polarity magnetic elements a) outside the green box, towards the left at (x,y)$\sim$(50,45), via long magnetic field lines and b) inside the green box, towards the right, (x,y)$\sim$(85,40), via short and low-lying magnetic field lines. The long magnetic field lines originate from the part of the network with the maximum magnetic flux density along the LOS, which, according to the SOT/SP magnetogram, reaches $\sim$1.2\,kG. The short, low-lying magnetic field lines are associated with closely packed \Ha absorption structures and appear darker in soft X-rays. The longer lines are partly outlined by radially distributed, elongated mottles and are associated with brighter coronal emission. 
\par

Figure~\ref{fig:rasters} shows the ROI in more detail, as imaged by the EIS and SOT/SP rasters. For comparison we have also included a ``synthetic'' \ha centre pseudo-raster, which was created as follows: after coaligning the \ha and EIS FOV, the temporal information of each $2\arcsec$ EIS slit exposure was used to select the closest, in time, $2\arcsec$-wide \ha slice along the Y-axis. \Ha shows the magnetized plasma of the chromosphere, while a comparison with the extrapolated magnetic field of the SOT/SP magnetogram indicates that dark mottles largely outline the magnetic field lines that originate at strong magnetic flux concentrations. 

Emission in the UV and EUV lines map the upper chromosphere, TR and corona. The \heii and \ov lines are associated with the network boundaries. The contrast between the network and the internetwork is lower in the \heii line than in the \ov line, in agreement with earlier observations \citep{judge_piet04}. The higher temperature \feviii and \sivii lines (as well as \mgvi and \mgvii) exhibit a different morphology. 

A comparison between the \feviii and \sivii emission and the extrapolated magnetic field shows that the TR plasma may be found either at the interface between the corona and the chromosphere, or in small scale low-lying loops, which do not reach coronal temperatures. From the following description, it will be evident, that these two ``types'' of transition region may be both associated with the same network boundary, depending on the magnetic connectivity of the region.

The left part of the network, associated with the longest magnetic field lines, exhibits very weak emission in \feviii and \sivii and reaches coronal temperatures (as inferred by the intense emission in \fexii). The right part of the network, associated with magnetic loops of various scales, shows intense emission in the \feviii and \sivii lines and weak or no coronal emission in Fe\,{\sc xii}. These low-lying loops appear also as absorption features in H$\alpha$. Intense emission in \feviii and \sivii accompanied by no coronal emmision in \fexii is also found at the lower part of the ROI, where the SOT/SP magnetogram shows numerous small scale magnetic concentrations with magnetic flux densities up to a few hundred Gauss. 

Therefore, the EUV TR emission is not explicitly associated to the expansion of the magnetic field of the network. At the same network boundary and depending on the magnetic connectivity of the region, the TR plasma may be found either at the interface between the corona and the chromosphere or in small scale low-lying loops which do not reach coronal temperatures. This is also illustrated at the RGB image in Figure~\ref{fig:rasters}h. \textit{Blue} represents the emission in \ov and marks the location of the network while \textit{red} is the overlying coronal emission of Fe\,{\sc xii}. The areas in \textit{green}, which represents \feviii, show us these two types of TR emission. 
\par
The first one, a ``classical'' TR, \textit{i.e.} a thin boundary between the chromospheric and coronal plasma is located around the network and follows the expansion of the network magnetic field lines. There, the average electron temperature and density (Figure~\ref{fig:rasters}f, \textit{green} rectangle) are equal to $5.74\pm0.08\cdot10^{5}\,K$ and $3.6\pm0.5\cdot10^{8}\,cm^{-3}$, correspondingly.
\par 
The second component, as already mentioned, is associated with magnetic flux tubes with very weak or no coronal counterpart at all. Some of them resemble the ``intra-network'' loops described by \citet{dowdy93} and connect different parts of the network. These are seen at (X,Y)$\sim$(80$\arcsec$,38$\arcsec$). Others are associated with field lines that connect weak magnetic elements at the IN, reach heights lower than 1\,Mm and can be as short as a few arcsec (lower right of the ROI). The \mgvii 278.39\,{\AA}/\mgvi 269.00\,{\AA} and \mgvii 280.75\,{\AA}/278.39\,{\AA} line ratios show us that the TR plasma in these small-scale loops is hotter and denser than the corresponding plasma found at the network.  At the ``intra-network'' loops (Figure~\ref{fig:rasters}f, \textit{red} rectangle) the corresponding values are $6.54\pm0.08\cdot10^{5}\,K$ and $5.2\pm0.5\cdot10^{8}\,cm^{-3}$ while at the low lying loops (Figure~\ref{fig:rasters}f, magenta rectangle) $6.37\pm0.08\cdot10^{5}\,K$ and $4.4\pm0.4\cdot10^{8}\,cm^{-3}$, respectively.

\begin{figure}[htp]
\centerline{ \includegraphics[width=1.05\textwidth]{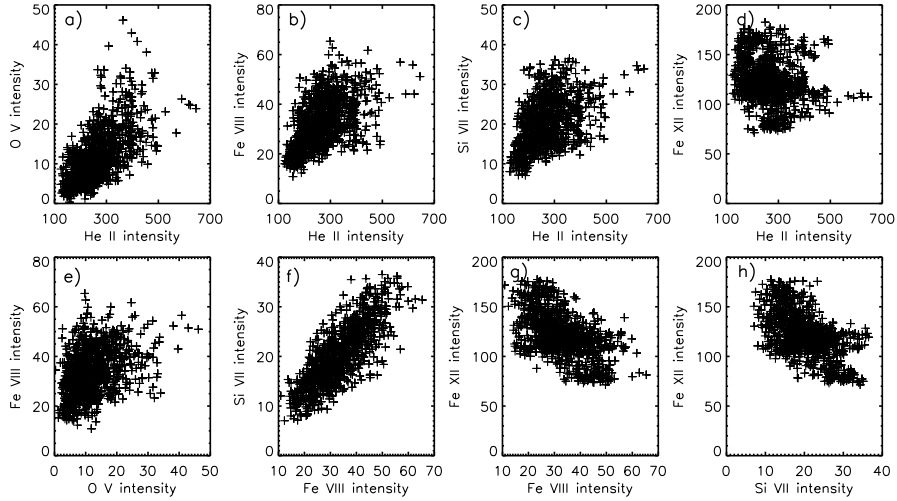}
 \vspace{0.01\textwidth}     }
  \caption{Scatter plots between the peak intensities of the EIS spectral lines of the rasters in Figure~\ref{fig:rasters}.}
    {\label{fig:int_scat}}
\end{figure}

\begin{figure}[htp]
\centerline{ \includegraphics[width=1.05\textwidth]{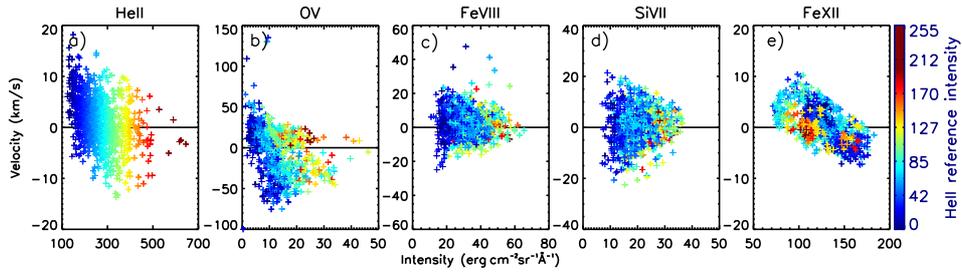}
 \vspace{0.05\textwidth}     }
  \caption{Velocity-Peak Intensity scatter plots for each of the spectral lines of the EIS raster in Figure~\ref{fig:rasters}. Different colors denote different \heii peak intensity values, scaled to the 0-255 range.}
    {\label{fig:param_scat}}
\end{figure}

The \heii velocity maps (Figure~\ref{fig:rasters}j) show some blueshifted features around the network. Some of these features coincide with (or appear to be the extensions of) areas with upward motions in H$\alpha$, but there is no pixel-by-pixel correlation between velocities. The \ov\ maps are very noisy and given the heavily blended line complex of the \caxvii window, it is a question how trustworthy these velocity measurements are. The velocity maps of \feviii and \sivii are also very noisy, but they show that the network is dominated by upward motions, which agrees with previous studies of spectral lines at these temperatures \citep{kayshap15}. In \fexii, the bright coronal structure is clearly associated with upflows. Lacking observations of the other end of the structure (which is located outside the ROI), we cannot conclude on the specifics of this flow (\textit{e.g.} if it is siphon flow or not).

\subsubsection{Intensity - intensity and intensity - velocity correlations}

The spatial structuring of the emission in different lines at the ROI as a function of temperature is also illustrated in the scatter plots of Figure~\ref{fig:int_scat}, where the \heii line intensity is used as a reference. The Pearson correlation coefficient between the peak intensity in the \heii and the other TR lines drops as the temperature of the latter increases (0.64, 0.49, 0.41, and -0.11 for \ov -- \heii, \feviii -- \heii, \sivii -- \heii and \fexii -- \heii, respectively). The lack of correlation between \fexii -- \heii shows that the coronal morphology of the network is altogether different from the chromospheric one. In Figure~\ref{fig:int_scat}e--h, we present scatter plots between peak intensities of different EIS lines. As expected, \ov emission is weakly correlated with that of Fe\,{\sc viii}. This result shows how emission in the upper TR originates mostly from the small-scale loops and to a lesser degree from the network itself. The high correlation between \sivii -- \feviii (0.8) indicates emission that comes from the same structures, while the moderate anticorrelation of these two lines with the coronal \fexii line (-0.55 and -0.57, Figure~\ref{fig:int_scat}g,h) shows that different parts of the network reach different temperatures. 
\par

In Figure~\ref{fig:param_scat} we plot the intensity-velocity relationship for the lines of the EIS raster shown in Figure~\ref{fig:rasters}. We use the \heii intensity as a reference. Different \heii (byte-scaled) intensity values are denoted with different colors and the transition from lower to higher \heii intensity represents the transition from the IN towards the network. As the \heii intensity increases, \heii velocity values shift from being predominantly positive (downflows) to being predominantly negative (upflows). We should note that for the full raster the \heii intensity-velocity scatter plot shifts from blue to red shifts, on average, for higher intensities, in accordance with other works \citep{brynildsen98}. The \ov line shows a similar transition but the velocity map is very noisy. A different velocity distribution arises at the upper TR lines (\feviii and \sivii), where velocities are roughly distributed symmetrically around zero. Above the network (corresponding to the highest \heii intensity), we find mostly upward motions. At the corona (\fexii line), the highest intensities are associated with blueshifts, which means that the bright coronal structures within the ROI is dominated by upward motions.

\par
In summary, the dominant structure in the ROI is a large funnel-like magnetic structure, rooted at the left part of the network region and containing plasma heated up to coronal temperatures. At the base of this structure there is a ``traditional'' TR, \textit{i.e.} a thin layer between the chromospheric and the coronal plasma. At the right part of the network, numerous shorter low-lying loops emanate, with various lengths down to a few arcsecs, which mostly contain plasma heated to upper TR temperatures and show no coronal emission. The funnel structure has $\sim$30$\%$ lower electron density than the low-lying loops, while the latter reach temperatures up to $6 \cdot 10^{5}$\,K. Furthermore, at a region outside the network, we find similar low-lying loops \textit{i.e.} reaching $6 \cdot 10^{5}$\,K and showing no coronal counterpart.

\begin{figure}[!ht]
\centerline{\includegraphics[width=1.05\textwidth]{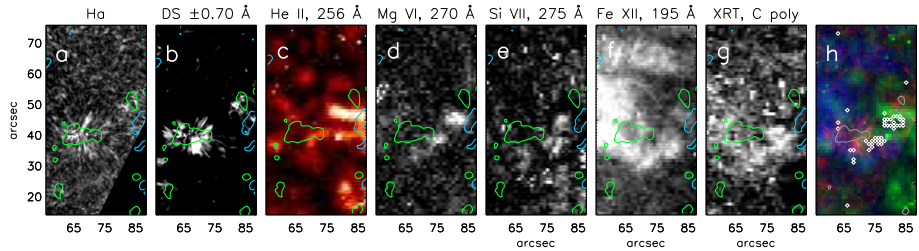}    }
  \caption{a--g) Standard deviation maps derived from the EIS slot time-series for the ROI. As a reference, \textit{green} (\textit{blue}) contours represent cotemporal positive (negative) magnetic flux density derived from the MDI magnetograms. h) RGB image showing DS$\pm$0.70\,\AA (\textit{red}), \heii\ (\textit{green}) and \fexii\ (\textit{blue}) standard deviation. Diamonds mark higher than 2$\sigma$ \mgvi variations while the grey contour outlines the magnetic footpoints of the network.}
    {\label{fig:sigmaps}}
\end{figure}

\subsection{EUV brightness variations and relation to \ha}
\label{s:euv_emission}

\par
We now turn to the analysis of the time-series of the slot images. In general, the slot images reveal the same morphology as Figure~\ref{fig:rasters}, but the corresponding features appear smoother since they are directly imaged and the EIS slot images mix spectral with spatial information. For this reason, the \feviii and \caxvii windows (which contains the \ov line) cannot be used. 
\par

In order to examine if there is any connection between the temporal variations at the chromosphere, TR and corona, we constructed the maps of standard deviation (``SDEV'' hereafter) of the 40$\arcsec$ slot intensities, \ha line center intensities and the DS at $\pm$0.70\,{\AA} from line center (Figure~\ref{fig:sigmaps}). 

High SDEV patches are found in all maps mostly around the magnetic field of the network. In the \ha intensity and DS maps (panels a--b), these patches are elongated and are associated to the chromospheric motions along the magnetic field lines while at the \heii, \mgvi and \sivii these are more roundish as per the lower spatial resolution of EIS. Also, the lower EIS sensitivity in the upper transition region emission \citep{young07} results to lower SDEV values and fewer patches. In the \fexii line and the XRT soft X-ray emission, high SDEV is found inside the bright coronal structure, around the network. We note that the temporal variations inside these patches are, in most cases, spatially coherent, showing that they are the result of a physical process and not noise. Most prominent is the elongated bright feature at (x,\,y)$\sim$(80,\,45), visible in the He\,{\sc ii}, Mg\,{\sc vi} and, to a lesser extent, in Si\,{\sc vii}. This patch corresponds to a continuous increase of EUV emission, due to a small scale emerging dipole. 

\begin{figure}[!ht]
\centerline{\includegraphics[width=1.\textwidth]{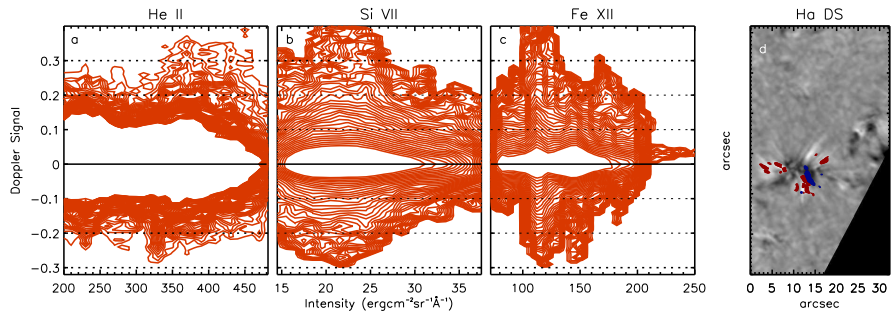}
     }
  \caption{a--c) Scatter plots between \ha DS $\pm$0.70\,{\AA} and He\,{\sc ii}, \sivii and \fexii intensities, respectively. For better comparison, \textit{horizontal lines} mark some DS values. Positive DS values denote upward motions. The density of points is plotted as \textit{contours} for better visibility. d) Average DS $\pm$0.70\,{\AA} map. Overplotted with \textit{red} are the locations where DS was higher than 0.3, and EUV intensities were simultaneously in the ranges 340--460 for He\,{\sc ii}, 19--26 for \sivii and 100--140 for Fe\,{\sc xii}. Similarly, in \textit{blue} we mark the locations where DS was lower than -0.2, and EUV intensities were simultaneously in the ranges 300--350 for He\,{\sc ii}, 15--28 for \sivii and 100--130 for Fe\,{\sc xii}}
    {\label{fig:euv_to_he}}
\end{figure}

The spatial distribution of high SDEV in all channels denotes the impact of the small scale events of the network region on the TR and the fanning-out of the magnetic field lines with height. We have plotted an RGB image containing information from all atmospheric layers in Figure~\ref{fig:sigmaps}h to illustrate this. Many \heii high SDEV patches (in \textit{green}) appear adjacent to or as extensions of the ones in \ha DS (in \textit{red}) while the coronal SDEV patches (in \textit{blue}) are located further outwards. For many of these features, there is also detectable temporal variation in the \mgvi and \sivii emission (\textit{white diamonds}). From the overall appearance of the RGB image in Figure~\ref{fig:sigmaps}h, we conclude that at least part of the emission variations in EIS windows, from the photosphere up to the corona, is related to the fine structures observed at the chromosphere. Although the resolution of EIS does not allow the detection of fine structures at the TR, as \textit{e.g.} in \citet{hansteen14}, our results confirm that in many cases, features at higher temperatures appear as extensions of chromospheric ones \citep{pereira14}. Through these EIS observations, this finding is now extended to higher TR and coronal temperatures.

By visual comparison of the \heii\ and \ha filtergrams, it may be coarsely implied that absorption features in \ha appear in emission in He\,{\sc ii}. The enhanced emission of the helium lines has lead researchers to propose alternative formation mechanisms for this line \citep[see \textit{e.g.}][for a comprehensive review]{judge_piet04}. Based on observations of He\,{\sc i} and \heii lines by SUMER and CDS, these authors suggest that the mixing of neutral helium with the coronal plasma, above the magnetic canopy, increases the emission in these lines. These ions can be advected either along or across the magnetic field and neutrals can be excited by the hot electrons found at higher atmospheric layers. As a result, the helium emission is overall enhanced and the contrast between network and internetwork regions is reduced. It is possible that macroscopic motions of chromospheric features, \textit{e.g.} spicules, may play a role in this process. It should be noted that \citet{andretta00} have found an anti-correlation between the \ha red wing and \heii intensities, while \citet{golding17}, showed that the enhanced \heii emission is caused primarily by non-equilibrium helium ionization effects.
\par
The effect of chromospheric motions along the magnetic field on the \heii emission can be inferred by the preceding description of the SDEV maps. High SDEV patches in \heii are the extensions of the corresponding chromospheric ones around the network (see Figure~\ref{fig:sigmaps}a, b and c). To further investigate the effect of chromospheric motions on the \heii emission we plot, in Figure~\ref{fig:euv_to_he}a, the \heii emission \textit{versus} the \ha DS at $\pm$0.70\,{\AA} from line center, for each pixel and time step. The distribution is largely symmetric, on top of which lie excursions of higher positive and negative DS. Highest positive values of the \ha DS at $\pm$0.70\,{\AA} (which correspond to upward chromospheric motion) are associated with \heii intensities higher than 340\,$erg\,cm^{-2}\,sr^{-1}\,{\AA}^{-1}$. Some of the downward motions also appear to be associated with enhanced \heii emission, but to a less extent. Taking into account that large DS at $\pm$0.70\,{\AA} may correspond to relatively higher Doppler shifts of the \ha profile, we conclude that the \heii intensity is affected by those \ha events with the highest velocities.
\par

We also examine whether there is an association between chromospheric motions and the emission in \sivii and \fexii\ slot data (Figure~\ref{fig:euv_to_he}b and ~\ref{fig:euv_to_he}c). For Si\,{\sc vii}, higher positive and negative DS produce emission enhancements up to 28\,$erg\,cm^{-2}\,sr^{-1}\,{\AA}^{-1}$. The effect of high positive and negative DS is also evident on coronal intensities. Overall, while upward motions produce highest emission in He\,{\sc ii}, the corresponding motions only contribute to low and intermediate emission in \sivii and Fe\,{\sc xii}. The locations where high \ha DS ($>$0.3) appear simultaneously with the corresponding intensity EUV enhancements were calculated. These are marked with red on the time-averaged DS map in Figure~\ref{fig:euv_to_he}d and are indeed located along features that exhibit persistent upward motion. Similarly, we over-plotted the enhancements that correspond to high negative \ha DS ($<$-0.2). These are found along a persistently red-shifted feature on the network. Taking into account projection effects (due to the slanted orientation of the small scale structures) which hinder a pixel by pixel comparison, the results shown in Figure~\ref{fig:euv_to_he} indicate that enhancements in the low/intermediate intensity range of upper atmospheric emission are mostly associated with upward chromospheric motions.

\subsection{Chromospheric jets and EUV emission}
\label{s:jets}

Regardless of the relatively low resolution of EIS, the \heii\ slot time-series reveals pronounced activity around the network. We present two examples of jets observed clearly in the \heii slots, that are related to \ha jet-like features and result to conspicuous variations at the overlying layers.

\begin{figure}[!ht]
\centerline{\includegraphics[width=1\textwidth]{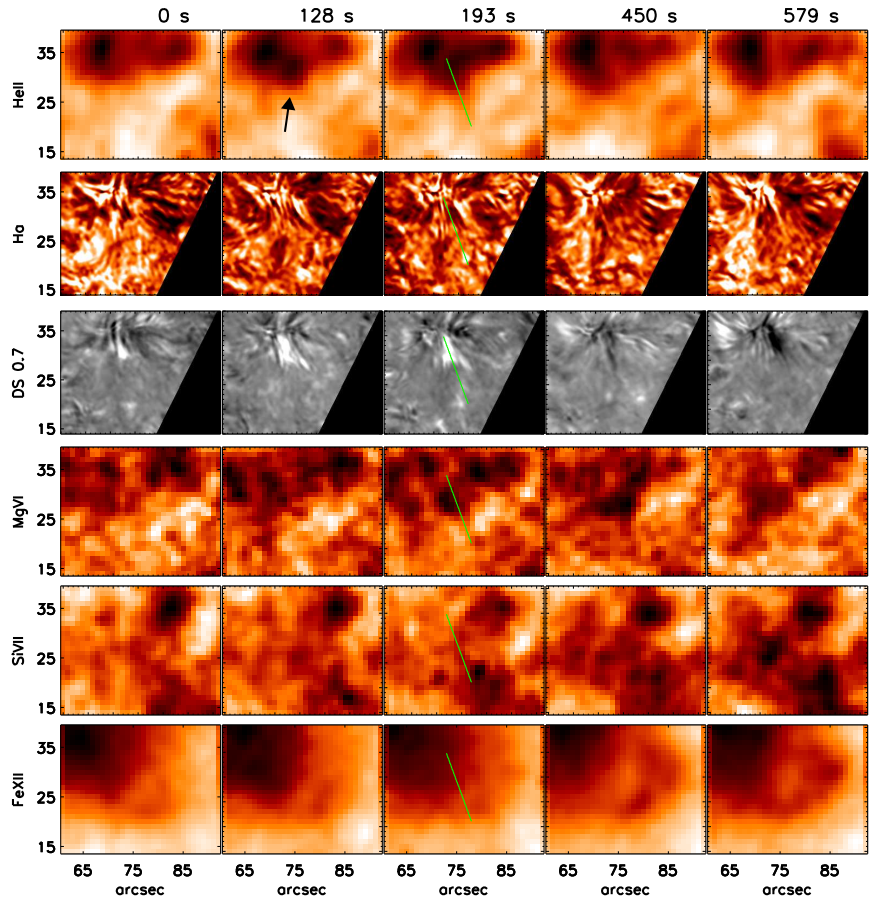}
 \vspace{0.07\textwidth}     }
  \caption{The evolution of an \heii brightening, which is cospatial with a chromospheric jet in H$\alpha$. Left column, from \textit{top} to \textit{bottom}: snapshots of \heii intensity, \ha centre intensity and \ha DS at $\pm$0.70\,{\AA}, Mg\,{\sc vi}, \sivii and Fe\,{\sc xii}. The EUV windows have been multiplied by a gaussian kernel and the color table has been inverted, for better visibility. The \textit{black arrow} at the top row indicates the position of the brightening in He\,II. The \textit{green} lines indicate the position where a slice was taken for the x-t diagram, shown in Figure~\ref{fig:jet1_slices}.}
    {\label{fig:jet1}}
\end{figure}

\begin{figure}[!ht]
\centerline{\includegraphics[width=0.7\textwidth]{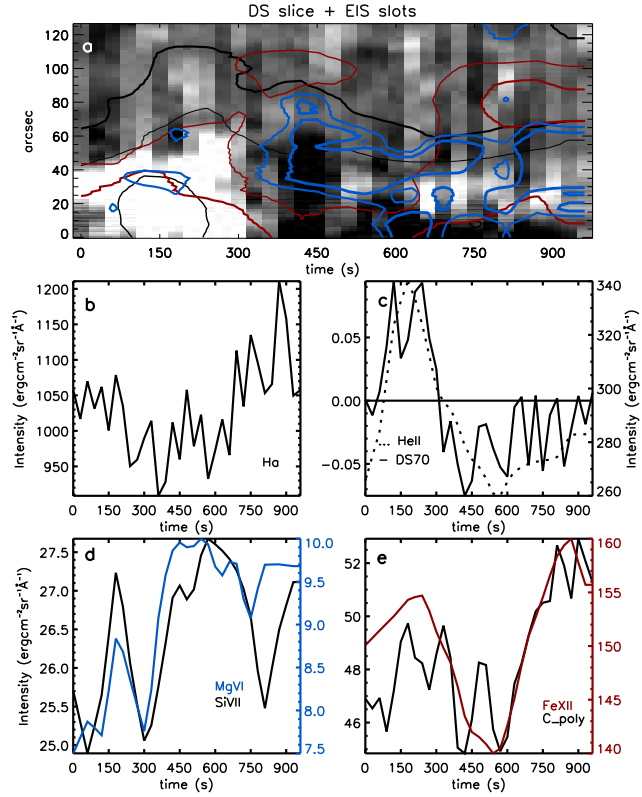}
 \vspace{0.07\textwidth}     }
  \caption{a) X-t diagram of the \ha DS at $\pm$0.70\,{\AA}, along the slice marked with the \textit{green} line in Figure~\ref{fig:jet1}, with over-plotted contours of \heii (black), \fexii (red) and \mgvi (blue). Thicker contours stand for more intense emission. b-e) summed intensities along the slices taken in \textit{different lines} (and \ha$\pm$0.7\,{\AA} DS).}
    {\label{fig:jet1_slices}}
\end{figure}

In Figure~\ref{fig:jet1} we present a TR brightening at (x,\,y)$\sim$(75,\,30), associated with a chromospheric jet (elongated white structure in \ha DS at $\pm$0.70\,{\AA}, third row). The increase in the \heii intensity (note that the brightest features appear dark in the EUV emission in Figure~\ref{fig:jet1}) is co-spatial with the upward moving jet and increased absorption from a bundle of mottles in the \ha (second and third row). The \ha jet has already started to evolve and the bright blob appears 1\,min later in \heii protruding from the network. Its expansion follows the evolution of the chromospheric jet.
In \heii it starts as a roundish brightening and evolves into a bifurcating jet-like feature.
In the next frame, a brightening also appears along the network, adjacent to the origin of the first blob. In \ha there is a bundle of mottles that exhibit intense upward motion. This upward moving chromospheric material has a similar bifurcated shape and, owing to the higher spatial resolution of DOT, we may discern several mottles that comprise the jet. The morphology of the feature in \ha and \heii is indicative of a three-dimensional structure seen at different heights. The jet is followed by downward chromospheric motions and then it fades from view in the \heii window. At the hotter \sivii window, no corresponding feature is detected, but there are variations in brightness, located further out from the jet. The \fexii slots show no elongated or blob-like features. However, related variations in intensity can be found in the x-t diagram (Figure~\ref{fig:jet1_slices}a, red contour) constructed along the structure (green line in Figure~\ref{fig:jet1}). The upward motion of the material corresponds to increased emission in both \heii and \fexii. Concerning the latter, this increase cannot be discerned against the overlying coronal emission in the images but is visible in the x-t diagram. Increased emission is detected at the \mgvi line which is co-spatial with the downflows in H$\alpha$ that follow the upward motion.

The lightcurves in Figure~\ref{fig:jet1_slices} were calculated by summing the signal along the entire slice (Figure~\ref{fig:jet1_slices}) for each time instance. It is worth noting that the event starts with upward motion of material seen in absorption in H$\alpha$. Its duration is about 7\,min during which, the intensity in \heii increases, following closely the DS curve (Figure~\ref{fig:jet1_slices}c). This evolution confirms the relation between \heii emission and upward chromospheric motion described in Sec.~\ref{s:euv_emission}. Soft X-ray and \fexii emission (Figure~\ref{fig:jet1_slices}e) both follow the DS curve very closely, demonstrating a coronal response to this jet. The rest of the TR lines also exhibit a small peak during the upward motion of the jet. Judging by the \ha DS curve, the material then descends and the intensity in \heii decreases. During this downward chromospheric motion, the intensity of the upper TR lines (\mgvi and Si\,{\sc vii}) increases notably (see the blue contour in Figure~\ref{fig:jet1_slices}a and d). This evolution suggests a connection between the chromospheric motion of material and emission up to the corona. Interestingly, the bulk of the TR emission during this event, is associated with the downflow. \citet{madjarska_doyle03} describe a similar (although larger) event which initiates as upflow and then becomes predominantly downflow, in the N\,{\sc v} 1238\,{\AA} line, while in \citet{nelson_doyle13}, a chromospheric jet exhibits signatures in the EUV channels up to the corona. It is possible that this TR brightening is associated with a macrospicule seen on disk \citep{madjarska06}.

\begin{figure}[htp]
\centerline{\includegraphics[width=1\textwidth]{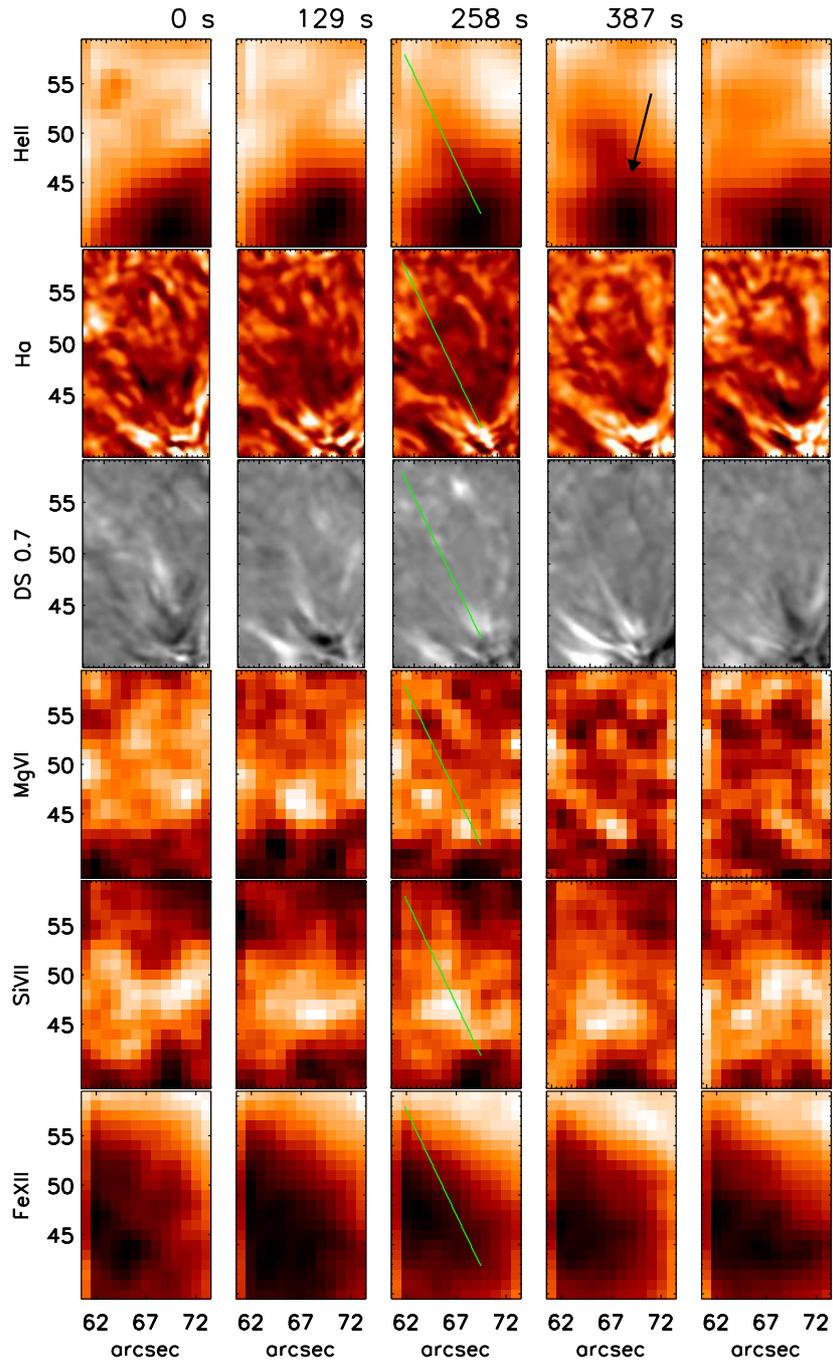}
 \vspace{0.07\textwidth}     }
  \caption{Same as Figure~\ref{fig:jet1} but for another jet-like feature.}
    {\label{fig:jet2}}
\end{figure}
\begin{figure}[htp]
\centerline{\includegraphics[width=0.7\textwidth]{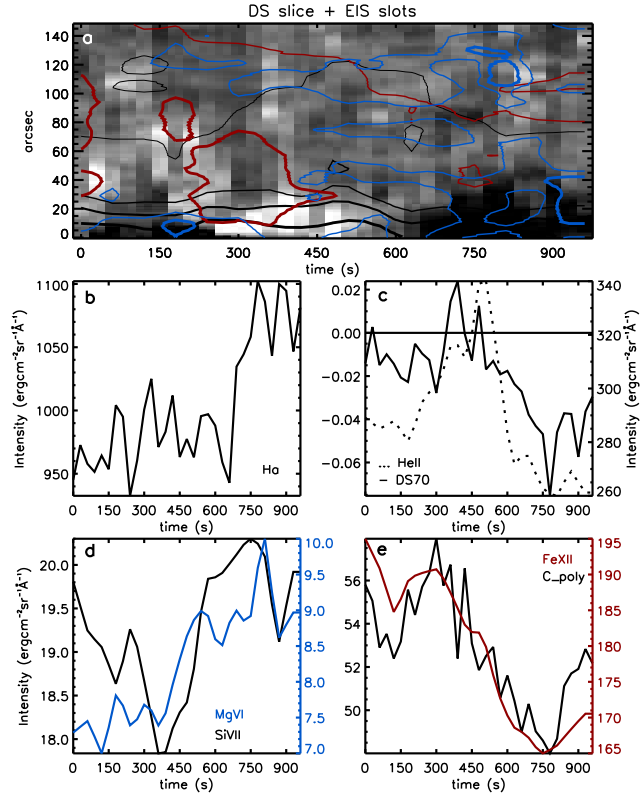}
 \vspace{0.07\textwidth}     }
  \caption{Same as Figure~\ref{fig:jet1_slices} but for the feature of Figure~\ref{fig:jet2}.}
    {\label{fig:jet2_slices}}
\end{figure}

The evolution of the jet in Figure~\ref{fig:jet1} suggests that the downflow of spicular (mottle) material could be the generator of TR region brightenings at the network \citep[see \textit{e.g.}][]{gallagher99,bewsher03}. The upper chromospheric (He\,{\sc ii}) and coronal (Fe\,{\sc xii}) emission are both co-temporal with the upward motion of the chromospheric material while the bulk of the TR emission follows a few minutes later, during the downflow of the material (negative \ha DS). Unfortunately, due to lack of spectral information for the TR lines, we do not know whether the emission in \mgvi and \sivii is accompanied by redshifts in these lines. Based on these observations, the origin of the high temperature component of the chromospheric jet, whose cooling seems to produce the subsequent increase in lower temperatures (TR lines) can only be conjectured. 
\par
Another example of a TR jet associated with a chromospheric feature is shown in Figure~\ref{fig:jet2}. This is more pronounced in the \heii emission. The \ha shows thin features with upward motion but shorter than the \heii feature. The DS maps (Figure~\ref{fig:jet2}) and x-t slice map (Figure~\ref{fig:jet2_slices}) show that at the footpoints of the jet, located at (x,\,y)$\sim$(70,\,43), there is clear upward motion (Figure~\ref{fig:jet2}, third row, third panel), preceded by a bi-directional flow (Figure~\ref{fig:jet2}, third row, second panel). In \ha mottles, these flows have been associated with heating and are attributed to reconnection \citep{tsirop_tzio04}. The \heii counterpart is more extended than the \ha feature and the \Ha absorption (seen also as lower intensity at Figure~\ref{fig:jet2_slices}b) is more intense near the footpoints. If a reconnection process is indeed producing this feature, then it is reasonable to expect that the plasma is heated, moving away from the formation temperature of \ha absorption and producing the peaks in \heii, \fexii and soft X-rays. These peaks are roughly co-temporal with the upward chromospheric motions at t$\sim$350\,s (Figure\ref{fig:jet2_slices}c and e). Then, emission at the TR lines starts to increase (Figure~\ref{fig:jet2_slices}d), after t$\sim$450\,s, when the coronal emission decreases and the chromospheric motion is downward, suggesting a further evolution similar to the previous case.
\par
If the observed bi-directional flow is attributed to a reconnection process, then a significant mass content can be driven towards upper atmospheric layers as suggested by \citet{tsirop_tzio04}. Unfortunately, the low resolution MDI data are not adequate to support this scenario. On the other hand, the high resolution SOT/SP data show that small scale magnetic fields are abundant at the IN so the interaction of small scale magnetic structures with the opposite polarity magnetic fields of the network provides support to a magnetic reconnection process as a realistic scenario.

\section{Conclusions and discussion}
\label{s:discussion}
In this work we use multi-wavelength observations of a quiet Sun region at the solar disk center, covering all atmospheric layers from the photosphere up to the corona. The observations consist of simultaneous H$\alpha$ line profile observations from DOT, TR and low corona observations from EIS/Hinode (both raster images with important spectral information and time-series of slot images) and soft X-ray filtergrams from XRT/Hinode  combined with magnetic field observations from MDI/SoHO and SOT/SP/Hinode. We exploit this unique dataset in order to examine the morphology, dynamics, temporal variability and relation to the magnetic field of the observed region, as well as the connectivity and interplay between the different atmospheric layers. 

As it is already stated \citep[see \textit{e.g.}][]{tian10} the magnetic structure of the transition region and corona and their connectivity with the chromosphere and photosphere are not completely understood.  Utilizing extrapolations of the observed magnetic field together with EUV observations we were able to show that along with the ``traditional'' TR, which acts as an interface between chromospheric and coronal plasma, significant emission in the TR temperatures comes from small-scale magnetic structures. These may be either part of the network or formed by the abundant small-scale magnetic fields of the IN. Their sizes are up to a few arcseconds, while signs of their emission at temperatures from 8$\cdot$10$^4$\,K to 6$\cdot$10$^5$\,K are found. This is a very important finding because it shows that numerous small-scale cool and of intermediate temperatures, low-lying loops are contributing to the TR emission. These results agree with the TR model of \citet{dowdy86} and the description of \citet{peter01} and \citet{schrijver03}. TR emission is mostly attributed to low-lying loops, but also associated with phenomena near the footpoints of coronal structures corroborating thus the results presented in \citet{patsourakos07} and \citet{tian10}.

Recent high resolution observations have shown that small-scale loops are abundant in EUV emission in temperatures near 10$^5$\,K \citep{guerreiro13,hansteen14,schmit16} but no comparison with high resolution magnetograms was made in these analyses. Small scale loops have been reported in active region plages in \citet{barczynski17}, but their association with magnetic fields was not feasible  due to the low resolution of the HMI magnetogram. The difficulty in quiet Sun is that the omnipresent small scale magnetic structures evade detection in low resolution magnetograms. Therefore, although it is reasonable to associate TR emission with the footpoints of coronal structures, the confirmed existence of small scale photospheric elements \citep{lites08} inevitably leads to the presence of short scale magnetic loops that connect the network boundaries with the IN, in a manner similar to what is described \textit{e.g.} in \citet{schrijver03}. It is also reasonable to expect that numerous small scale magnetic loops will permeate the internetwork. The high resolution SOT/SP magnetogram and the calculated magnetic field lines in Figure~\ref{fig:rasters} now justify this assumption, since small-scale magnetic elements are abundant in the FOV and are connected with short field lines. As seen at the lower right of the panels in Figure~\ref{fig:rasters}, these small-scale structures are associated with emission up to the \sivii formation temperature and very weak or no coronal emission. 

Previous studies have utilized co-temporal EUV wide-slit observations by CDS and have shown that there is a correspondence between chromospheric and coronal temporal variations \citep{ugarte04}. Here, we extend these studies by examining simultaneous high resolution chromospheric \ha observations along with higher resolution TR and coronal observations for the first time. We showed that upper TR emission variations are located mostly around the network and appear as extensions of or co-spatially with the underlying chromospheric variations. The SDEV maps of Figure~\ref{fig:sigmaps} allow the conjecture that chromospheric mottles may reach temperatures at least up to 6$\cdot$10$^5$\,K. It has already been shown that some chromospheric features appear more extended in higher TR temperatures observed by IRIS \citep{pereira14}. We extend this finding to even higher TR temperatures, in the spectral lines observed by EIS. 

Two individual jets related to the network and stand out clearly in H$\alpha$ observations were studied. It was found that brightenings in the TR and corona may be the result of typical chromospheric motions, not necessarily the fastest ones. In these events, the bulk of the TR emission at the vicinity of the funnel-like structure is associated with the downflows that follow. We were also able to examine, for the first time, the impact of chromospheric motions, as derived through the H$\alpha$ Doppler signals (a far more reliable measure of chromospheric motions than the intensity on one position on the H$\alpha$ profile used in some studies), on the transition region and coronal emission recorded with the slot images of EIS. The highest chromospheric DS seem to be related to increased emission in TR temperatures. In the case of \heii, the highest emission is associated with large H$\alpha$ blueshifts, but it is relatively unaffected by redshifts, justifying a velocity redistribution mechanism \citep{macpherson_jordan99,andretta00}. In the case of \sivii and \fexii emission, we find an association between Doppler-shifted chromospheric events and low-to-intermediate EUV intensities. These associations are found along persistent upward-moving chromospheric features. It is quite interesting that not necessarily the fastest events (which are usually the focus of other studies), but rather typical network structures are associated with these intensity enhancements. 

Recent high resolution observations have shown that highly Doppler-shifted events appear at the chromosphere in very short time-scales \citep{rouppe09}. So far, the difference in the resolution achieved by instruments that provide chromospheric and upper TR/coronal observations has not facilitated easy comparison between features at different atmospheric heights and temperature regimes. To overcome this difficulty, \citet{mcintosh_depontieu09a} have studied line asymmetries in EUV spectra, concluding that they point to a heating mechanism during rapidly-changing chromospheric events. In the recent works of \citet{rouppe15} and \citet{henriques16}, it is shown that, indeed, rapidly-changing chromospheric events are associated with brightenings up to coronal temperatures. In addition, observations in plages and coronal holes near the limb have shown that these rapidly-changing events are associated with coronal heating \citep{depontieu11}. Although there is an ongoing debate on whether these events may indeed produce heating \citep[see \textit{e.g.}][]{klimchuk12,klimchuk14}, our analysis contributes to the existing (and increasing) literature in a twofold manner: 1) by showing that there is an EUV counterpart of Doppler-shifted chromospheric features at the upper TR and coronal temperature emission, even at the quiet solar conditions and 2) that not necessarily the fastest of these events are associated with this emission. These findings were deduced from a very quiet region, at the solar disk center with no criterion on the H$\alpha$ DS imposed on the observations; besides, the spatial, temporal and spectral resolution of DOT does not allow for the detection of very fast moving features at the chromosphere. If indeed these events are related with rapid variations in EUV temperatures, then, arguably, the low temporal-spatial resolution of EIS smears them out and they appear as low magnitude intensity enhancements. 
\par
Our results demonstrate the potential of observations like the ones provided by the EIS instrument. The ability to record emission in several spectral windows, although at the expense of spectral resolution, is invaluable, because it can provide simultaneous overview of plasmas at different temperatures, especially when combined with high resolution magnetograms and chromospheric observations. Of course, the chromospheric events described in Sec.~\ref{s:jets} are of small scale. However, the corresponding signals in the EUV, although extremely weak, are conspicuous. This is an indication that TR is highly variable even at the smallest scales and we believe that a combination of high quality images and spectral observations will shed light on the relations between chromospheric fine structures, TR transients, and coronal emission.

\begin{acknowledgements}
The authors would like to thank the anonymous referee whose valuable comments greatly improved the manuscript.
The observations have been funded by the Optical Infrared Coordination network (OPTICON, http://www.ing.iac.es/opticon), a major international collaboration supported by the Research Infrastructures Program of the European Commission''s sixth Framework Program. The research was partly funded through the project ``SOLAR-4068',' which is implemented under the ``ARISTEIA II'' Action of the  operational program ``Education and Lifelong Learning'' and is cofunded by the European Social Fund (ESF) and Greek national funds. The DOT was operated at the Spanish Observatorio del Roque de los Muchachos of the Instituto de Astrof\'{i}sica de Canarias. The authors thank P. S\"{u}tterlin for the DOT observations and R. Rutten for the data reduction. Hinode is a Japanese mission
developed and launched by ISAS/JAXA, collaborating with NAOJ as a domestic
partner, and NASA and STFC (UK) as international partners. Scientific operation
of the Hinode mission is conducted by the Hinode science team organized at
ISAS/JAXA. This team mainly consists of scientists from institutes in the partner countries. Support for the post-launch operation is provided by JAXA and
NAOJ (Japan), STFC (U.K.), NASA, ESA, and NSC (Norway). Hinode SOT/SP
Inversions were conducted at NCAR under the framework of the Community
Spectro-polarimetric Analysis Center (CSAC; http://www.csac.hao.ucar.edu). The authors would like to thank Dr S.H. Park for valuable help in the magnetic field extrapolation.
\end{acknowledgements}

\bibliographystyle{spr-mp-sola}
\bibliography{references}

\end{article}
\end{document}